\def\year{2023}\relax
\documentclass[letterpaper]{article} % DO NOT CHANGE THIS
\usepackage{aaai23}  % DO NOT CHANGE THIS
\usepackage{times}  % DO NOT CHANGE THIS
\usepackage{helvet}  % DO NOT CHANGE THIS
\usepackage{courier}  % DO NOT CHANGE THIS
\usepackage[hyphens]{url}  % DO NOT CHANGE THIS
\usepackage{graphicx} % DO NOT CHANGE THIS
\usepackage{natbib}  % DO NOT CHANGE THIS AND DO NOT ADD ANY OPTIONS TO IT
\usepackage{caption} % DO NOT CHANGE THIS AND DO NOT ADD ANY OPTIONS TO IT
\DeclareCaptionStyle{ruled}{labelfont=normalfont,labelsep=colon,strut=off} % DO NOT CHANGE THIS
\usepackage{algorithm}
\usepackage{algorithmic}
\usepackage{blindtext}
\usepackage[utf8]{inputenc}
\DeclareUnicodeCharacter{2009}{\,}

\usepackage{newfloat}
\usepackage{listings}
\floatstyle{ruled}
\newfloat{listing}{tb}{lst}{}
\floatname{listing}{Listing}

\title{What We Know So Far: Artificial Intelligence in African Healthcare}
\author{
    Naome Etori,   %\thanks{With help from the AAAI Publications Committee.}\\
    Ebasa Temesgen, and
    Maria Gini \\
}
\affiliations{
    Department of Computer Science and Engineering\\
    University of Minnesota\\
    200 Union Street SE Minneapolis, MN 55455 \\
    \{etori001,temes021, gini\}@umn.edu
}

\usepackage{bibentry}
\begin{document}

\maketitle

\begin{abstract}

Healthcare in Africa is a complex issue influenced by many factors including poverty, lack of infrastructure, and inadequate funding. However, Artificial intelligence (AI) applied to healthcare, has the potential to transform healthcare in Africa by improving the accuracy and efficiency of diagnosis, enabling earlier detection of diseases, and supporting the delivery of personalized medicine. This paper reviews the current state of how AI Algorithms can be used to improve diagnostics, treatment, and diseases monitoring, as well as how AI can be used to improve access to healthcare in Africa as a low-resource setting and discusses some of the critical challenges and opportunities for its adoption. As such, there is a need for a well-coordinated effort by the governments, private sector, healthcare providers, and international organizations to create sustainable AI solutions that meet the unique needs of the African healthcare system.
\end{abstract}

\section{Introduction}
The application of AI in healthcare dates back to the 1950s. Researchers at the Massachusetts Institute of Technology (MIT) developed ``Project MAC'' a program to help analyze and interpret medical data, one of the first examples of AI in healthcare. MAC stood for ``machine-aided cognition'' \cite{garfinkel1999architects}.

In the 1960s, AI was used to create knowledge-based expert systems, that mimic the decision-making processes of human experts. The ``MYCIN'' system, developed at Stanford University to aid in diagnosing and treating infectious diseases, was an early example of an expert system in healthcare \cite{feigenbaum1981expert}.

During the 1980s and 1990s, AI-based Machine learning algorithms could learn and adapt to new data without being explicitly programmed. This led to the development of programs that could analyze medical images, such as X-rays, to help diagnose diseases and AI algorithms used for diagnosing, identifying, and analyzing public health threats \cite{kaul2020history}.

More recently, the improvements in  natural language processing (NLP) has been used to understand and interpret human language. NLP techniques have been applied in healthcare to extract and analyze data such as electronic health records (EHRs),medical images and assist with tasks (such as appointment scheduling and medication management). As a result of increases in computing power, the availability of large amounts of data, and the rapid development of big data analytics have enabled AI applications. Deep learning (DL) machine learning technology, which involves training artificial neural networks (ANN) on large datasets, has been a major driver of recent AI advances. This has significantly impacted modernizing the global healthcare system to improve diagnosis and clinical care accuracy and efficiency. A convolutional neural network (CNN) is a type of DL algorithm that simulates the behavior of interconnected neurons in the human brain \cite{kaul2020history}.

The recent surge in AI healthcare research in Africa indicates the potential of AI to improve patient outcomes and reduce the burden on the healthcare system. This paper discusses AI algorithms (such as expert systems, machine learning, deep learning, natural language processing, and image processing) and how they are applied in the African healthcare systems, including challenges in the application of AI systems in Africa as well as opportunities for AI adoption. 

\section{Problem Definition}
There has been a lot of research on AI in healthcare, but there has been minimal discussion about AI in African healthcare. AI researchers have also found that there is a lack of diverse representation of AI models in healthcare and particularly from low-resource languages. To fill this gap, we address two main research questions:

\textbf{RQ1:} \textit{ What role do AI algorithms play in African healthcare systems?} 

\textbf{RQ2:} \textit{How does a lack of resources impact AI implementation in Africa?} 

Our goal was to understand and explain the current integration and implementation of AI in African healthcare systems to inspire future AI research in African Healthcare.

\section{Related Work}
Many researchers have shown that AI has the potential to improve patient care and lower healthcare costs. Our review of the literature yielded the following topics for discussion.

\subsection{Recent AI Advances in Healthcare in Developed Countries }
AI has made significant advancements in the healthcare industry in developed countries in recent years. This is because AI algorithms can handle and learn from vast amount of healthcare data generated due to advancement of AI technologies. Deep learning, in particular has impacted how we view AI tools today and is the source of the recent advancement in AI applications \cite{bohr2020rise}.

Sophisticated AI algorithms have been applied in diagnosis and treatment planning to analyze medical images, such as X-rays and CT scans, to help doctors diagnose diseases \cite{liu2020deep,vaishya2020artificial,lin2019diagnostic,hu2018deep} and decide plan treatment. For example, \cite{lakshmanaprabu2019optimal,holbrook2021detection,makaju2018lung,ardila2019end} shows how an AI algorithm was able to detect lung cancer from CT scans with a very high degree of accuracy, comparable to that of an experienced radiologist. 

Big data has created opportunities for AI-based Machine learning (ML) algorithms, enabling them to learn from previous data to make accurate predictions about new data. This can assist healthcare providers in prioritizing their resources and intervening before a patient condition deteriorates\cite{senders2018machine,stafford2020systematic,lee2018applications,choi2016doctor}. 

Pharmaceutical companies use AI algorithms to identify new candidates for drug development and analyze the effects of different drugs \cite{david2020molecular,paul2021artificial,smith2018transforming,smalley2017ai,fleming2018artificial,chan2019advancing,lavecchia2019deep}.

AI-powered Virtual assistants help triage patients and provide them with information about their condition and treatment options \cite{bharti2020medbot,bates2019health,fadhil2018beyond}

The use of clinical decision support \cite{loftus2020artificial,lysaght2019ai} and medication management \cite{howe2006perioperative} have improved patient care while reducing the workload for healthcare professionals in hospitals.

\subsection{Recent AI Advances in African Healthcare}
AI has enormous potential to help in the prediction and prevention of outbreaks of infectious diseases. Data from the World Health Organization (WHO) \cite{nkiruka2021prediction} shows the relationship between malaria incidence and various climate variables, found that specific patterns in the data were indicative of increased malaria risk, and used machine learning algorithms to create a model for predicting malaria incidence based on these patterns. Research conducted by \cite{colubri2016transforming} used past Ebola outbreak data to train the ML algorithm to accurately predict the outcomes of Ebola patients with a high degree of accuracy which could be used in future outbreaks to help guide clinical decision-making and improve patient outcomes.

AI can improve the accuracy of diagnoses and treatment recommendations in Africa. Due to a lack of resources for optimal healthcare provision, trained physicians, and underfunded public healthcare facilities, ML models can be used to analyze medical images and provide diagnoses and chronic diseases such as cancer. \cite{adeoye2022artificial} shows the application and implementation phases of oncological AI tools in Africa using patient cohorts in Africa.  

\subsection{Status of AI in African Healthcare system}
The application of AI in the African healthcare system is still in its early stages. However, there is a growing interest and investment in the implementation of AI to improve different aspects of healthcare delivery in Africa. However, the implementation of AI in Africa confronts numerous challenges\cite{owoyemi2020artificial}. Many African countries, for example, still have limited access to reliable electricity and to high-speed internet, making it challenging to implement and use AI systems. The successful implementation of AI in health care depends on the availability and dependability of the infrastructure\cite{mbunge2022we}. 

For many years, African healthcare systems have suffered from man-made issues such as institutional, human resource, financial, technical, and political developments \cite{oleribe2019identifying}. Hence, most African countries cannot meet the fundamental requirements for effective healthcare systems. Ineffective service integration is linked to poor governance, and human resource challenges \cite{nutbeam2000health,marais2015health}. The lack of basic access to healthcare makes it even harder to implement AI solutions  \cite{tran2019global}.

Additionally, the scarcity of trained professionals with expertise in AI in Africa is a major concern; hence, many healthcare systems are struggling to meet the rising demand for services while also facing significant shortages of trained health workers and essential medicines \cite{naicker2009shortage,guo2018application}. As a result, most African countries rely heavily on developed nations for AI technologies to solve critical healthcare challenges due to inadequate funding \cite{guo2018application}. 

Ethical considerations around the use of AI in healthcare in Africa, including issues related to data privacy and the potential for biased decision-making could contribute to slow AI implementation in the African region \cite{hagerty2019global}.

\subsection{AI Gaps in Africa: Key Challenges}
African healthcare systems are lagging in implementing AI  mainly due to lack of resources. Implementing AI in healthcare can be expensive, and securing funding for such initiatives can be challenging in most African countries. Ensuring that AI models are sustainable in the long term is also an important consideration \cite{wakunuma2020socio}. 

\begin{itemize}
\item There is often a lack of high-quality data available for training machine learning models due to poor record-keeping and inadequate infrastructure for data collection and storage. For example, doctors and nurses still use hand written notes when seeing patients. This can make it difficult to develop accurate and effective AI systems for the African region. Prior studies have demonstrated disparate AI performance by race AI models, especially on African datasets. The training datasets may not be representative of the African population. If the datasets used to train the models are not diverse or representative enough, the models may not be able to generalize well to new, unseen data from Africa population \cite{banerjee2021reading,kamulegeya2019using,harmon2020artificial,obermeyer2019dissecting}.

\item Furthermore, the AI models may not have been specifically designed to handle the unique challenges present in African datasets. For example, African datasets may contain a greater variety of languages, dialects, and accents, and the models may not have been trained on data that reflects this linguistic diversity. In addition, there may be other cultural and societal factors that are unique to Africa that the models have not been designed to handle. According to estimates, 17\% of the world's languages, many of which are spoken in Africa, are "low-resource languages" in the digital realm \cite{gwagwa2020artificial}.

\item Despite these challenges, there have been some efforts to improve healthcare in Africa. For example, some African governments have increased funding for healthcare, and there have been efforts to train more healthcare professionals and build new healthcare facilities, as many developing countries are attempting to move more towards universal healthcare coverage \cite{lagomarsino2012moving}. In addition, some international organizations and charities have provided assistance to boost the delivery of healthcare in Africa \cite{gibbs2014evolution,beigbeder1991role}.

\end{itemize}

\section{Methodology}
This paper reviews the current state of AI in healthcare in Africa, including the challenges and opportunities for its adoption. A comprehensive search of the literature was conducted using several databases. The survey reviewed 30 journal papers obtained electronically through four scientific databases (Google Scholar, Scopus, IEEE, Pum Med, and Science Direct) searched using three sets of keywords: (1) Artificial Intelligence in Africa (2) Artificial Intelligence in Healthcare (3) African Healthcare Systems. We limited our search to articles that made algorithmic contributions and addressed diseases in their applications. We disqualified papers that only provided opinions, theories, surveys, or datasets. This yielded a total of 12 papers.

\section{Findings}
\subsection{RQ1: Role of AI Algorithms on African Healthcare }
Based on our findings, different algorithms have been used to address different healthcare,  such as Expert systems, machine learning, and deep learning. Although  expert systems have been extensively researched globally, little research has been conducted on its application in African healthcare systems. Expert systems can be beneficial, especially where scarcity of trained physicians and medical personnel exists, as is usually the case in African countries. Expert systems can help African healthcare systems diagnose patients and select treatment plans without extensively trained medical personnel, where a decision must be made quickly to save lives \cite{wahl2018artificial}. An example is where an expert system has been incorporated with fuzzy logic systems, to improve the diagnosis of chronic conditions like STDs, HIV/AIDS, cholera, abdominal pain, and diabetes decision support application in South Africa \cite{6428572, thompson2017expert, fathi1994medusa, fleming2007fuzzy, lee2010fuzzy}. 

Natural language processing (NLP) has been used in developing a Medical Chatbot to diagnose patients in their early stages of the disease or to use social media data for surveillance and monitoring of infectious disease outbreaks \cite{gupta2020social, abdelwahap2021applications}. 
The use of NLP in African healthcare is still at its infancy stage. \cite{li2020we} conducted mental health condition research following the outbreak of coronavirus(COVID-19), using Twitter data from Nigeria and South Africa. \cite{oyebode2018likita} demonstrated that Likita, a chatbot, could be used to diagnose common ailments and improve healthcare delivery in Africa.

Deep learning (DL) can process large amounts of data, such as images, and could potentially aid medical workers in decision-making, using an X-ray image to analyze multiple diseases. For instance, \cite{stephen2019efficient} used X-ray images to classify Pneumonia with a validation accuracy of 93.73\%. \cite{simi2022early} also used X-ray data to diagnose early stage Tuberculosis using a DL algorithm and achieved 99\% accuracy. \cite{holbrook2021detection} used three pre-trained DL models (faster R-CNN, single-shot multi-box detector (SSD), and RetinaNet) for the microscopic diagnosis of malaria parasites, malaria is one of the deadliest diseases in Sub-Sahara Africa in thick blood smears. The result found that a faster R-CNN has a higher accuracy than the other two models used in the study with an average precision of over 0.94. Hence this approach can improve the accuracy and efficiency of malaria diagnosis compared to traditional methods.

ML models are used to predict and classify chronic diseases in Africa. They are gaining popularity for their simplicity and the availability of data in a few African countries. Table 1 has summarized some of the research on applying AI to address different diseases in  African countries.

\subsection{RQ2: How does the lack of resources impact AI implementation in Africa?}
African healthcare systems suffer from the lack of data access due to inadequate resources, which is often a key ingredient in the development and training of AI systems. Lack of infrastructure and investment in data collection and storage has made it difficult for organizations to obtain the data they need to develop and train AI systems \cite{banerjee2021reading,kamulegeya2019using,harmon2020artificial,obermeyer2019dissecting}

%\subsection{Deep Learning}
%Deep learning (DL) is used to process large amounts of data, such as images, and could potentially aid medical workers in decision-making. Using an X-ray image to analyze multiple diseases. For instance, \cite{stephen2019efficient} used X-ray images to classify Pneumonia with a validation accuracy of 93.73\%. \cite{simi2022early} also used X-ray data to diagnose Tuberculosis at an early stage using a Deep learning algorithm and achieved 99\% accuracy. 

%\subsection{Machine Learning}
%Its simplicity of application and the availability of data in a few African countries has increased the research on Machine learning. Table 1 has summarized some of the research that has been conducted in the application of AI to address different types of diseases in different African countries.   

\begin{table*}[!htbp]
\caption{Application of AI and Machine Learning in Africa healthcare System}
\label{tab:table}
\begin{tabular}
{| p{0.1\linewidth}|p{0.15\linewidth}|p{0.15\linewidth} | p{0.15\linewidth} | p{0.15\linewidth} | p{0.16\linewidth} |}

\hline
\textbf{Disease} & \textbf{Country} & \textbf{Algorithm} & \textbf{Dataset} & \textbf{Accuracy} & \textbf{Reference} 

\\
\hline
HIV/AIDS & S.Saharan.A & XGBoost  &  PHIA & 90\% males, 92\% female & \cite{mutai2021use} \\
\hline
HIV/AIDS & Kenya, Uganda  &  - &Data from 16 communities  & 78\% & \cite{balzer2020machine} \\
\hline
Malaria & Burkina Faso  &  Random forest &georeferenced raster(SPOT)&84\% & \cite{taconet2021data} \\
\hline
Ebola & W.Africa & Central Bayesian & 842 molecules& 83\% & \cite{anantpadma2019ebola}
\\
\hline

Colorectal Cancer & S.Africa & Artificial NN & WDGMC CRC & 87.0\% & \cite{achilonu2021predicting}\\
\hline
Breast cancer & Nigeria & Random Forest &Surgical data(LASUTH) &  96.67\%  & \cite{macaulay2021breast} \\
\hline
Cervical cancer & African countries & Ensemble learning &  SMOTE & 87.21\% & \cite{ahishakiye2020prediction}\\ 

\hline
COVID-19 & African Countries& Ensemble  &  (WHO) database & MAD = 0.0073, MSE = 0.0002, R2 = 0.9616 & (Ibrahim et al. 2023)\\
\hline
Diabetes & Nigeria &  Binarized Naïve Bayes
(BNB) & Collected(nairaland) &  87.08\% & \cite{oyebode2019detecting}\\
\hline
Coronary Heart D & South Africa &  Naïve Bayes,SVM and Decision Tree & S.Africa-KEEL & higher than 70\% & 
 \cite{gonsalves2019prediction}\\ 
 
\hline
Obstetric Fistula & Tanzania & Logistic Regression& Collected from
CCBRT,Dar-es-Salaam & 86\% & (Fihavango et al. 2021) \\

\hline
Malaria & Sub-Saharan & Decision Tree & (WHO) datasets & 75.10\% & \cite{masinde2020africa} \\
\hline

\end{tabular}
\end{table*}

AI in African healthcare has generally been used in diseases mapping such as HIV. ML techniques identify HIV predictors for screening in sub-Saharan Africa \cite{mutai2021use}, malaria predictions \cite{nkiruka2021prediction}. Even though African healthcare systems strive to implement AI to assist with healthcare operations, it is still difficult to respond to public health emergencies such as disease outbreaks, resulting in increased mortality and morbidity.

\section{Discussion}
AI has been applied in various healthcare settings in Africa,  as shown in Table \ref{tab:table}. Hence has the potential to significantly improve Human Immunodeficiency Virus (HIV) care in Africa by providing more accurate diagnoses, predicting patient outcomes, and optimizing treatment programs. For example, \cite{bauer2019global} used DL to improve the accuracy of HIV diagnoses in sub-Saharan Africa, \cite{mutai2021use} applied ML approaches for building models in identifying HIV predictors as well as predicting persons at high risk of infection, the XGBoost algorithm significantly improved the identification of HIV positivity by f1 scoring mean of 90 and 92\% for males and females. respectively. Also, \cite{balzer2020machine} used  ML to Identify Persons at High-Risk of HIV Acquisition in rural Kenya and Uganda, and the result shows that ML improved efficiency by 78\%. 

Malaria has been (and still is) a major public health crisis in Africa, with an estimated 435 million cases and 1.3 million deaths yearly (WHO, 2021). Experts say malaria slows economic growth in Africa by up to 1.3 percent per year. AI can be used in the fight against malaria in Africa; for example, \cite{taconet2021data} used data-driven and interpretable ML modeling Algorithm (Random Forest) to explore malaria vector biting rates in rural Burkina Faso. The results identified several aspects of the bio-ecology of the main malaria vectors \cite{masinde2020africa} Decision trees algorithm and climate data produced accuracy results of 99\%.

Furthermore, AI has the potential to significantly improve the response to Ebola outbreaks in Africa by providing more accurate and timely predictions, faster diagnoses, and more efficient use of resources; for example, Bayesian (ML) Models were used to enable new in Vitro Leads\cite{anantpadma2019ebola,lane2019natural}.

Research has highlighted the potential benefits of using AI to improve cancer diagnosis and treatment in Africa. For example, in predicting colorectal cancer recurrence and patient survival using Artificial neural network(ANN), a supervised ML approach where scored the highest AUC-ROC for recurrence (87\%) and survival (82\%) in South Africa \cite{achilonu2021predicting}. Random forest (ML) models have been used in Breast cancer risk prediction among African women in Nigeria by \cite{macaulay2021breast}; the Chi-Square selected features gave the best performance with 98.33\% accuracy, 100\% sensitivity, 96.55\% specificity, and 98\% AUC. Also,\cite{ahishakiye2020prediction} predicted Cervical Cancer Based on risk factors using the Ensemble Learning model with an accuracy of 87.21\%.

Since the COVID-19 pandemic, the demand for use of AI in the African healthcare sector to aid in the pandemic response has increased. Many African countries have been conducting studies on this area; for example, \cite{ibrahim2022multi} conducted a study to predict daily COVID-19 cases spreading across the north, south, east, west, and central Africa regions and countries using  ML, including Artificial neural network(ANN), adaptive neuro-fuzzy inference system (ANFIS), support vector machine (SVM) and convent and the result shows high accuracy performance. 

According to the International Diabetes Federation (IDF), 24 million adults (ages 20-79) have diabetes, which accounts for 416,000 deaths in the IDF Africa Region in 2021. AI can potentially improve diabetes care in Africa by improving access to care and providing more personalized treatment options. For example, using social media and ML  \cite{oyebode2019detecting} detected factors responsible for Diabetes prevalence in Nigeria.

Management and treatment of coronary heart disease (CHD) in African healthcare systems have been a huge challenge. Recently, there has been a  growing recognition of the potential for AI to improve CHD disease treatment and care. For example, by using historical medical and ML models such as Naïve Bayes, SVM, and Decision Tree (DT) data to predict CHD  \cite{gonsalves2019prediction}.

WHO estimated more than 2 million young women live with untreated obstetric fistula (OF) in Asia and sub-Saharan Africa. Limited research on the use of AI in obstetric fistula disease management and treatment in Africa has been seen. For example, \cite{fihavango2021using} used Data Mining techniques to Predict Obstetric Fistula in Tanzania.

This survey did not cover the entire African continent. The results are promising, and the future of AI tools and algorithms to improve healthcare in the African continent look promising.

\section{Conclusions}
AI has the potential to predict and control diseases, expand and augment service delivery, and address several lingering social inequities by improving the accuracy and efficiency of diagnosis, enabling earlier detection of diseases, and supporting the delivery of personalized medicine. However, several challenges must be addressed to realize its full potential, including the lack of infrastructure, limited access to data, and the need for regulatory frameworks. African countries must build the necessary infrastructure to support technological advancements. This can include investing in high-speed internet, data centers, and cyber security systems. Governments can help promote the development and adoption of AI by establishing clear AI regulatory standards. Private sector, healthcare providers, partnership with international stakeholders can design a sustainable AI solutions that meet the unique needs of the African healthcare system

\section{Future Work}
AI and machine learning are critical to addressing health care inadequacies in African countries. AI models may perform poorly on African datasets, However, to improve the performance of AI models on African datasets, it will be necessary to create a more diverse and representative datasets as well as design models that are specifically tailored to deal with the unique challenges presented in African data.

% Use instead or the References section will not appear in your paper
\bibliography{aaai23}
% \section{Acknowledgments}
% AAAI is especially grateful to Peter Patel Schneider for his work in implementing the original aaai.sty file, liberally using the ideas of other style hackers, including Barbara Beeton. We also acknowledge with thanks the work of George Ferguson for his guide to using the style and BibTeX files.

\end{document}